# An algorithm for dividing quaternions

Alexandr CARIOW, Galina CARIOWA

West Pomeranian University of Technology, Szczecin, Faculty of Computer Science and Information Technology, Żołnierska 49, Szczecin 71-210, Poland

**Abstract:**

*In this work, a rationalized algorithm for calculating the quotient of two quaternions is presented which reduces the number of underlying real multiplications. Hardware for fast multiplication is much more expensive than hardware for fast addition. Therefore, reducing the number of multiplications in VLSI processor design is usually a desirable task. The performing of a quaternion division using the naive method takes 16 multiplications, 15 additions, 4 squarings and 4 divisions of real numbers while the proposed algorithm can compute the same result in only 8 multiplications (or multipliers – in hardware implementation case), 31 additions, 4 squaring and 4 division of real numbers.*

## 1. Introduction

Currently, quaternions [1] are widely used for data processing in various fields of science and engineering including physics [2], robotics [3,4], autonomous vehicles [5, 6], digital signal and image processing [7–14], computer graphics and machine vision [15, 17], wireless communications [18], public-key cryptography [19], neural networks [20-22], etc.

In quaternion algebra, the most time and area consuming operations are multiplication and division of two quaternions, what is more, the division is even more expensive than multiplication. The schoolbook multiplication of quaternions requires performing 16 real multiplications and 12 real additions, and the schoolbook division of quaternions requires performing 16 real multiplications, 4 real squaring, 15 real additions and 4 real divisions. In turn, multiplication and division of two ordinary real numbers are also more time-consuming operations than addition or subtraction of ordinary real numbers. Since these operations are carried out repeatedly, the total time to the correct implementation of the final algorithm may be unacceptable. It is therefore evident that finding ways, which reduce the number of real multiplications required for performing multiplication and division of quaternions, is a very important task. Efficient algorithms for multiplication of quaternions exist [23–25]. No such algorithms for the division of quaternions have been proposed. The aim of the present paper is to suggest an efficient algorithm for this purpose.

## 2. Preliminary remarks

A quaternion is a hypercomplex number which has can be presented as a linear combination:

$$q = q_0 + iq_1 + jq_2 + kq_3,$$

where $\{q_i\}, i = 0,1,2,3$ are real numbers, and $i$, $j$, and $k$ are imaginary units whose products are defined by the following table [1]:

| × | 1 | $i$ | $j$ | $k$ |
|---|---|---|---|---|
| 1 | 1 | $i$ | $j$ | $k$ |
| $i$ | $i$ | $-1$ | $k$ | $-j$ |
| $j$ | $j$ | $-k$ | $-1$ | $i$ |
| $k$ | $k$ | $j$ | $-i$ | $-1$ |

Suppose we are given two quaternions $q = q_0 + iq_1 + jq_2 + kq_3$ and $r = r_0 + ir_1 + jr_2 + kr_3$. Assume that we wish to divide $q$ by $r$. In other words, we need to calculate the quotient of two quaternions:

$$g = \frac{q}{r} = y_0 + iy_1 + jy_2 + ky_3.$$

A schoolbook method of finding the quotient of two quaternions (left division) can be represented by the following equations[*]:

$$y_0 = \frac{r_0 q_0 + r_1 q_1 + r_2 q_2 + r_3 q_3}{r_0^2 + r_1^2 + r_2^2 + r_3^2}, \quad y_1 = \frac{r_0 q_1 - r_1 q_0 - r_2 q_3 + r_3 q_2}{r_0^2 + r_1^2 + r_2^2 + r_3^2},$$

$$y_2 = \frac{r_0 q_2 + r_1 q_3 - r_2 q_0 - r_3 q_1}{r_0^2 + r_1^2 + r_2^2 + r_3^2}, \quad y_3 = \frac{r_0 q_3 - r_1 q_2 + r_2 q_1 - r_3 q_0}{r_0^2 + r_1^2 + r_2^2 + r_3^2}.$$

Direct implementation of the calculations in accordance with these equations requires 16 multiplications, 15 additions, 4 squarings and 4 divisions of real numbers. Below we will show how you can reduce the computational complexity of the operation of calculating the quotient of two quaternions.

## 3. The algorithm

Let us introduce the two column vectors: $\mathbf{X}_{4\times 1} = [r_0, r_1, r_2, r_3]^T$, $\mathbf{Y}_{4\times 1} = [y_0, y_1, y_2, y_3]^T$.

Let also $R = r_0^2 + r_1^2 + r_2^2 + r_3^2$.

---

[*] https://www.mathworks.com/help/aeroblks/quaterniondivision.html

Now the quaternion division can be presented in the matrix–vector multiplication form as

$$\mathbf{Y}_{4\times 1} = \frac{1}{R}\mathbf{Q}_4 \mathbf{X}_{4\times 1} \qquad (1)$$

where

$$\mathbf{Q}_4 = \begin{bmatrix} q_0 & q_1 & q_2 & q_3 \\ q_1 & -q_0 & -q_3 & q_2 \\ q_2 & q_3 & -q_0 & -q_1 \\ q_3 & -q_2 & q_1 & -q_0 \end{bmatrix}.$$

The direct multiplication of the vector–matrix product in Eq. (1) requires 16 real multiplications and 12 additions. We propose some tricks, which reduce the multiplicative complexity of this operation to 8 real multiplications at the price of 16 more real additions.

Let us multiply by (-1) the first row of the matrix and then multiply by (-1) the second, third and fourth columns of the obtained matrix. We can easily see that this transformation leads in the future to minimize the computational complexity of the final algorithm.

As a result, we obtain the following matrix:

$$\breve{\mathbf{Q}}_4 = \begin{bmatrix} -q_0 & q_1 & q_2 & q_3 \\ q_1 & q_0 & q_3 & -q_2 \\ q_2 & -q_3 & q_0 & q_1 \\ q_3 & q_2 & -q_1 & q_0 \end{bmatrix}.$$

The idea is that the matrix $\breve{\mathbf{Q}}_4$ can be decomposed as an algebraic sum of a symmetric Toeplitz matrix and another matrix which has many zero entries. Symmetric Toeplitz matrices are precisely those matrices that can be diagonalized by the Discrete Hadamard Transform – a transform which has a fast algorithm for its matrix-vector multiplication. This fact we will use in our further arguments.

So, we can write

$$\mathbf{Y}_{4\times 1} = \frac{1}{R}\mathbf{R}_4^{(2)}(\breve{\mathbf{Q}}_4 - 2\widehat{\mathbf{Q}}_4)\mathbf{R}_4^{(1)}\mathbf{X}_{4\times 1} \qquad (2)$$

where

$$\breve{\mathbf{Q}}_4 = \begin{bmatrix} q_0 & q_1 & q_2 & q_3 \\ q_1 & q_0 & q_3 & q_2 \\ q_2 & q_3 & q_0 & q_1 \\ q_3 & q_2 & q_1 & q_0 \end{bmatrix}, \quad \widehat{\mathbf{Q}}_4 = \begin{bmatrix} q_0 & 0 & 0 & 0 \\ 0 & 0 & 0 & q_2 \\ 0 & q_3 & 0 & 0 \\ 0 & 0 & q_1 & 0 \end{bmatrix},$$

$$\mathbf{R}_4^{(1)} = diag(1,-1,-1,-1), \quad \mathbf{R}_4^{(2)} = diag(-1,1,1,1).$$

Indeed, it is easy to see that $\breve{\mathbf{Q}}_4$ has the following structure:

$$\breve{\mathbf{Q}}_4 = \begin{bmatrix} \breve{\mathbf{Q}}_2^{(0)} & \breve{\mathbf{Q}}_2^{(1)} \\ \breve{\mathbf{Q}}_2^{(1)} & \breve{\mathbf{Q}}_2^{(0)} \end{bmatrix},$$

where

$$\breve{\mathbf{Q}}_2^{(0)} = \begin{bmatrix} q_0 & q_1 \\ q_1 & q_0 \end{bmatrix}, \; \breve{\mathbf{Q}}_2^{(1)} = \begin{bmatrix} q_2 & q_3 \\ q_3 & q_2 \end{bmatrix}.$$

It is easily verified [26] that the matrix with this structure can be effectively factorized than the computational procedure for calculating the product of $\breve{\mathbf{Q}}_4 \mathbf{R}_4^{(1)} \mathbf{X}_{4\times 1}$ can be represented as follows:

$$\breve{\mathbf{Q}}_4 \mathbf{R}_4^{(1)} \mathbf{X}_{4\times 1} = \mathbf{W}_4^{(0)} \mathbf{D}_4^{(0)} \mathbf{W}_4^{(0)} \mathbf{R}_4^{(1)} \mathbf{X}_{4\times 1} \qquad (3)$$

where

$$\mathbf{W}_4^{(0)} = \mathbf{H}_2 \otimes \mathbf{I}_2, \; \mathbf{D}_4^{(0)} = \frac{1}{2}[(\breve{\mathbf{Q}}_2^{(0)} + \breve{\mathbf{Q}}_2^{(1)}) \oplus (\breve{\mathbf{Q}}_2^{(0)} - \breve{\mathbf{Q}}_2^{(1)})],$$

$$\breve{\mathbf{Q}}_2^{(0)} + \breve{\mathbf{Q}}_2^{(1)} = \begin{bmatrix} q_0 + q_2 & q_1 + q_3 \\ q_1 + q_3 & q_0 + q_2 \end{bmatrix}, \breve{\mathbf{Q}}_2^{(0)} - \breve{\mathbf{Q}}_2^{(1)} = \begin{bmatrix} q_0 - q_2 & q_1 - q_3 \\ q_1 - q_3 & q_0 - q_2 \end{bmatrix}.$$

$\mathbf{H}_2 = \begin{bmatrix} 1 & 1 \\ 1 & -1 \end{bmatrix}$ - (2×2) Hadamard matrix, $\mathbf{I}_N$ - is order $N$ identity matrix, and „$\otimes$", „$\oplus$" – denote tensor product and direct sum of two matrices respectively [26, 27].

It is easy to see that the matrices $\breve{\mathbf{Q}}_2^{(0)} + \breve{\mathbf{Q}}_2^{(1)}$ and $\breve{\mathbf{Q}}_2^{(0)} - \breve{\mathbf{Q}}_2^{(1)}$ possess a structure that can get suitable decompositions too. We can therefore write:

$$\breve{\mathbf{Q}}_2^{(0)} + \breve{\mathbf{Q}}_2^{(1)} = \mathbf{H}_2 \mathbf{D}_2^{(0)} \mathbf{H}_2, \; \breve{\mathbf{Q}}_2^{(0)} - \breve{\mathbf{Q}}_2^{(1)} = \mathbf{H}_2 \mathbf{D}_2^{(1)} \mathbf{H}_2,$$

$$\mathbf{D}_2^{(0)} = \frac{1}{2}\{[(q_0 + q_2) + (q_1 + q_3)] \oplus [(q_0 + q_2) - (q_1 + q_3)]\},$$

$$\mathbf{D}_2^{(1)} = \frac{1}{2}\{[(q_0 - q_2) + (q_1 - q_3)] \oplus [(q_0 - q_2) - (q_1 - q_3)]\}.$$

Combining partial decompositions in a single procedure we can rewrite (3) as follows:

$$\breve{\mathbf{Q}}_4 \mathbf{R}_4^{(1)} \mathbf{X}_{4\times 1} = \mathbf{W}_4^{(0)} \mathbf{W}_4^{(1)} \mathbf{D}_4^{(1)} \mathbf{W}_4^{(1)} \mathbf{W}_4^{(0)} \mathbf{R}_4^{(1)} \mathbf{X}_{4\times 1}$$

where

$$\mathbf{W}_4^{(1)} = \mathbf{I}_2 \otimes \mathbf{H}_2, \; \mathbf{D}_4^{(1)} = diag(s_0, s_1, s_2, s_3)$$

$$s_0 = \frac{1}{4}[(q_0 + q_2) + (q_1 + q_3)], \; s_1 = \frac{1}{4}[(q_0 + q_2) - (q_1 + q_3)],$$

$$s_2 = \frac{1}{4}[(q_0 - q_2) + (q_1 - q_3)], \ s_3 = \frac{1}{4}[(q_0 - q_2) - (q_1 - q_3)].$$

It is easy to see that the entries of the matrix $\mathbf{D}_4^{(1)}$ can be calculated using the following vector–matrix procedure:

$$\mathbf{S}_{4\times 1} = \frac{1}{4} \mathbf{W}_4^{(1)} \mathbf{W}_4^{(0)} \mathbf{Q}_{4\times 1},$$

where

$$\mathbf{S}_{4\times 1} = [s_0, s_1, s_2, s_3]^\mathrm{T}, \ \mathbf{Q}_{4\times 1} = [q_0, q_1, q_2, q_3]^\mathrm{T}.$$

The computational complexity of the product $2\widehat{\mathbf{Q}}_4 \mathbf{R}_4^{(1)} \mathbf{X}_{4\times 1}$ cannot be reduced and this product is calculated directly, without any tricks. Combining the calculations for both matrices in a single procedure we finally obtain:

$$\mathbf{Y}_{4\times 1} = \mathbf{\eta}_4 \mathbf{\Sigma}_{4\times 8} \mathbf{W}_8^{(0)} \mathbf{W}_8^{(1)} \mathbf{D}_8 \mathbf{W}_8^{(1)} \widetilde{\mathbf{W}}_8^{(0)} \widetilde{\mathbf{P}}_{8\times 4} \mathbf{X}_{4\times 1}$$

where

$$\mathbf{D}_8 = diag(s_0, s_1, s_2, s_3, s_4, s_5, s_6, s_7), \ \mathbf{\eta}_4 = \eta \mathbf{R}_4^{(2)}, \ \eta = 1/R,$$

$$s_4 = 2q_0, \ s_5 = 2q_1, \ s_6 = 2q_2 \ s_7 = 2q_3,$$

$$\widetilde{\mathbf{W}}_8^{(0)} = (\mathbf{H}_2 \otimes \mathbf{I}_2) \oplus \mathbf{P}_4^{(0)} = \left[\begin{array}{cccc:cccc} 1 & 0 & 1 & 0 & 0 & 0 & 0 & 0 \\ 0 & 1 & 0 & 1 & 0 & 0 & 0 & 0 \\ 1 & 0 & -1 & 0 & 0 & 0 & 0 & 0 \\ 0 & 1 & 0 & -1 & 0 & 0 & 0 & 0 \\ \hdashline 0 & 0 & 0 & 0 & 1 & 0 & 0 & 0 \\ 0 & 0 & 0 & 0 & 0 & 0 & 0 & 1 \\ 0 & 0 & 0 & 0 & 0 & 1 & 0 & 0 \\ 0 & 0 & 0 & 0 & 0 & 0 & 1 & 0 \end{array}\right],$$

$$\mathbf{W}_8^{(0)} = (\mathbf{H}_2 \otimes \mathbf{I}_2) \oplus \mathbf{I}_4 = \left[\begin{array}{cccc:cccc} 1 & 0 & 1 & 0 & 0 & 0 & 0 & 0 \\ 0 & 1 & 0 & 1 & 0 & 0 & 0 & 0 \\ 1 & 0 & -1 & 0 & 0 & 0 & 0 & 0 \\ 0 & 1 & 0 & -1 & 0 & 0 & 0 & 0 \\ \hdashline 0 & 0 & 0 & 0 & 1 & 0 & 0 & 0 \\ 0 & 0 & 0 & 0 & 0 & 1 & 0 & 0 \\ 0 & 0 & 0 & 0 & 0 & 0 & 1 & 0 \\ 0 & 0 & 0 & 0 & 0 & 0 & 0 & 1 \end{array}\right],$$

$$\boldsymbol{\Sigma}_{4\times 8} = \overline{\mathbf{1}}_{1\times 2} \otimes \mathbf{I}_4 = \begin{bmatrix} 1 & 0 & 0 & 0 & -1 & 0 & 0 & 0 \\ 0 & 1 & 0 & 0 & 0 & -1 & 0 & 0 \\ 0 & 0 & 1 & 0 & 0 & 0 & -1 & 0 \\ 0 & 0 & 0 & 1 & 0 & 0 & 0 & -1 \end{bmatrix}, \mathbf{P}_4^{(0)} = \begin{bmatrix} 1 & & & \\ & & 1 & \\ & 1 & & \\ & & & 1 \end{bmatrix}, \overline{\mathbf{1}}_{1\times 2} = [1,-1],$$

$$\tilde{\mathbf{P}}_{8\times 4} = \mathbf{1}_{2\times 1} \otimes \mathbf{I}_4 \mathbf{R}_4^{(1)} = \begin{bmatrix} 1 & 0 & 0 & 0 \\ 0 & -1 & 0 & 0 \\ 0 & 0 & -1 & 0 \\ 0 & 0 & 0 & -1 \\ \hdashline 1 & 0 & 0 & 0 \\ 0 & -1 & 0 & 0 \\ 0 & 0 & -1 & 0 \\ 0 & 0 & 0 & -1 \end{bmatrix}, \mathbf{W}_8^{(1)} = (\mathbf{I}_2 \otimes \mathbf{H}_2) \oplus \mathbf{I}_4 = \begin{bmatrix} 1 & 1 & 0 & 0 & 0 & 0 & 0 & 0 \\ 1 & -1 & 0 & 0 & 0 & 0 & 0 & 0 \\ 0 & 0 & 1 & 1 & 0 & 0 & 0 & 0 \\ 0 & 0 & 1 & -1 & 0 & 0 & 0 & 0 \\ \hdashline 0 & 0 & 0 & 0 & 1 & 0 & 0 & 0 \\ 0 & 0 & 0 & 0 & 0 & 1 & 0 & 0 \\ 0 & 0 & 0 & 0 & 0 & 0 & 1 & 0 \\ 0 & 0 & 0 & 0 & 0 & 0 & 0 & 1 \end{bmatrix},$$

Indeed, it is easy to see, that the entries of the matrix $\mathbf{D}_8$ can be calculated using the following vector–matrix procedure:

$$\mathbf{S}_{8\times 1} = \tilde{\mathbf{D}}_8 \mathbf{W}_8^{(1)} \tilde{\mathbf{W}}_8^{(1)} \mathbf{P}_{8\times 4} \mathbf{Q}_{4\times 1} \tag{6}$$

where

$$\mathbf{S}_{8\times 1} = [s_0, s_1, s_2, s_3, s_4, s_5, s_6, s_7]^T, \tilde{\mathbf{D}}_8 = diag(1/4, 1/4, 1/4, 1/4, 2, 2, 2, 2),$$

$$\tilde{\mathbf{W}}_8^{(1)} = (\mathbf{H}_2 \otimes \mathbf{I}_2) \oplus \mathbf{P}_4^{(1)} = \begin{bmatrix} 1 & 0 & 1 & 0 & 0 & 0 & 0 & 0 \\ 0 & 1 & 0 & 1 & 0 & 0 & 0 & 0 \\ 1 & 0 & -1 & 0 & 0 & 0 & 0 & 0 \\ 0 & 1 & 0 & -1 & 0 & 0 & 0 & 0 \\ \hdashline 0 & 0 & 0 & 0 & 1 & 0 & 0 & 0 \\ 0 & 0 & 0 & 0 & 0 & 0 & 1 & 0 \\ 0 & 0 & 0 & 0 & 0 & 0 & 0 & 1 \\ 0 & 0 & 0 & 0 & 0 & 1 & 0 & 0 \end{bmatrix}, \mathbf{P}_4^{(1)} = \begin{bmatrix} 1 & & & \\ & 1 & & \\ & & & 1 \\ & 1 & & \end{bmatrix},$$

$$\mathbf{P}_{8\times 4} = \mathbf{1}_{2\times 1} \otimes \mathbf{I}_4 = \begin{bmatrix} 1 & & & \\ & 1 & & \\ & & 1 & \\ & & & 1 \\ \hdashline 1 & & & \\ & 1 & & \\ & & 1 & \\ & & & 1 \end{bmatrix}, \mathbf{1}_{2\times 1} = [1,1]^T.$$

Fig. 1 shows a data flow diagram of the rationalized algorithm for computation of the quotient of two quaternions and Fig. 2 shows a data flow diagram of the process for calculating the matrix $\mathbf{D}_8$ entries. In this paper, data flow diagrams are oriented from left to right. Straight lines in the figures denote the operations of data transfer. Points, where lines converge, denote summation. The dashed lines indicate the sign change operation. We deliberately use the usual lines without arrows on purpose, so as not to clutter the picture. The circles in these figures show the operation of multiplication by a number (variable or constant) inscribed inside a circle. In turn, the rectangles indicate the matrix-vector multiplications with the $(2 \times 2)$-Hadamard matrices.

## 4. Computation complexity

And now we have to count how many real multiplications (excluding multiplications by the power of two) and additions are required and compare this with the number required for a direct evaluation of matrix–vector product in Eq. (1). We can see that direct realization requires 16 multiplications, 15 additions, 4 squarings and 4 divisions of real numbers. The number of real multiplications required using our proposed algorithm is 8. Thus using the proposed algorithm the number of real multiplications to implement the quaternion division is reduced. The number of real additions required using our algorithm is 31. We observe that the direct computation of quotient requires 16 additions less than the proposed algorithm. Thus, our proposed algorithm saves 8 multiplications but increases 16 additions compared with the direct method. The number of divisions and squaring remained on the same level as in the naive method.

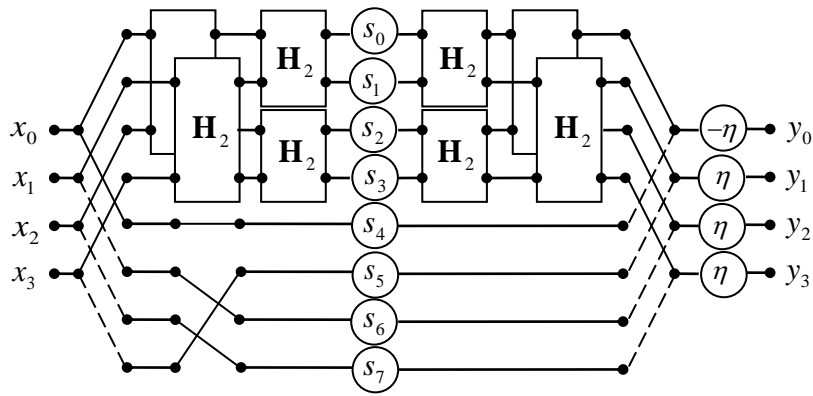

Fig. 1. Data flow diagram for rationalized quaternion division algorithm.

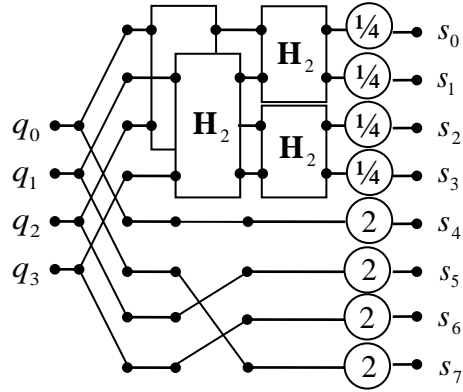

Fig. 2. The data flow diagram describing the process of calculating entries of the matrix $\mathbf{D}_8$ in accordance with the procedure (6).

Unfortunately, the algorithm additionally requires 8 multiplications by numbers that are powers of two. However, multiplication by powers of two is reduced to trivial shifts of the binary representations of the operands corresponding to the right or left. Such operations are very simple, and their execution time is usually neglected when assessing the complexity of computations.

## 5. Conclusion

We presented a new algorithm for calculating the quotient of two quaternions. The use of this algorithm reduces the multiplication complexity of quaternion dividing, thus reducing hardware complexity and leading to a high-speed architecture suitable for VLSI implementation. It is clear that the multiplication requires much more intensive hardware resources than addition. Hardware multiplier occupies much more area and consumes much more power than an adder. This is because the hardware complexity of a multiplier grows quadratically with operand size, while the hardware complexity of an adder increases linearly with operand size. Therefore, from point of view of VLSI implementation, an algorithm for dividing two quaternions containing as few real multiplications as possible is preferred. This is especially important in the case of a fully parallel implementation of the quaternion divider when each real multiplication is implemented by its own hardware multiplier.